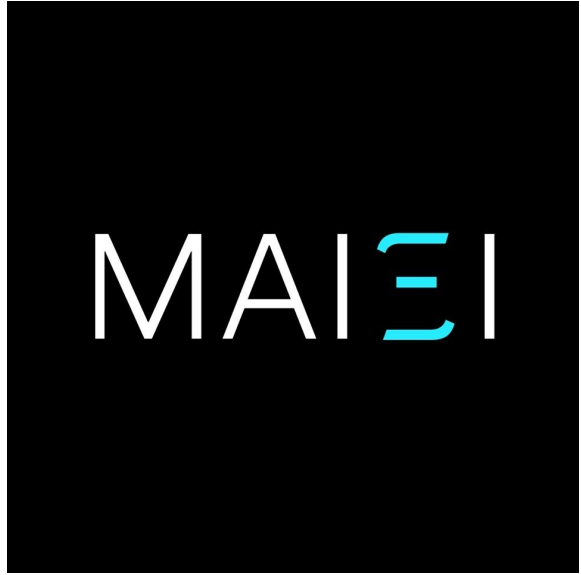

Montreal AI Ethics Institute

*An international, non-profit research institute helping humanity define its place in a world increasingly driven and characterized by algorithms*

Website: https://montrealethics.ai
Newsletter: https://aiethics.substack.com

# Report prepared by the Montreal AI Ethics Institute for Scotland's AI Strategy

*Based on insights and analysis by the Montreal AI Ethics Institute (MAIEI) Staff on the policy document from Scotland Government and supplemented by workshop contributions from the AI Ethics community convened by MAIEI on May 4, 2020*

Primary contact for the report:

Abhishek Gupta (abhishek@montrealethics.ai)
Founder, Montreal AI Ethics Institute
Machine Learning Engineer, Microsoft

## What do you think of the proposed definition of AI for the purposes of the strategy?

The proposed definition captures the essence of the meaning of Artificial Intelligence (AI) and while there isn't widespread consensus on its definition across the various groups that use the phrase, we propose adding a subtext to the definition to capture the fluidity and nuance of AI. Specifically, the term AI has over time shifted significantly in what it means and there are differences in perception of the term by the wider public compared to the intended definition as put forth by an organization. For example, the work from Lipton and Steinhardt[1] highlights how terms such as AI have social implications which can create widening gaps between reality and expectation because of concepts like *suggestive definitions* which communicate ability that means various things to the public which are different from the actual capabilities of the system. So when talking about concepts like visual perception, speech recognition, and decision-making, there is a certain expectation that is set that gives the impression that these systems exhibit human-like intelligence characteristics when in fact it has been shown in various forms that the representations captured by these systems are in fact quite brittle, vulnerable to attacks[2], and don't necessarily demonstrate intelligence in the humanistic sense where it is general enough to translate from one domain to another without significant retraining and re-tooling of the system, even in the case of narrow domains[3].

Moreover, from a policy and regulation perspective, when utilizing a definition that conveys human-like intelligence implications leads to problems of anthropomorphization of these technologies[4] which can lead to over-regulation and a chilling effect on investments and development of the field. This problem is exacerbated by problems of misrepresentations of the capabilities of the system[5], both intentional and unintentional, by journalists, companies, researchers, and others for various purposes which leads the public astray sparking unwarranted fears and discussions[6] while distracting from more meaningful and immediate concerns that need to be considered when crafting a national strategy for promoting responsible, safe, and ethical development of AI.

---

[1] Lipton, Z. C., & Steinhardt, J. (2018). Troubling trends in machine learning scholarship. arXiv preprint arXiv:1807.03341.

[2] Papernot, N., McDaniel, P., Goodfellow, I., Jha, S., Celik, Z. B., & Swami, A. (2017, April). Practical black-box attacks against machine learning. In Proceedings of the 2017 ACM on Asia conference on computer and communications security (pp. 506-519).

[3] Raffel, C., Shazeer, N., Roberts, A., Lee, K., Narang, S., Matena, M., ... & Liu, P. J. (2019). Exploring the limits of transfer learning with a unified text-to-text transformer. arXiv preprint arXiv:1910.10683.

[4] Proudfoot, D. (2011). Anthropomorphism and AI: Turing's much misunderstood imitation game. Artificial Intelligence, 175(5-6), 950-957.

[5] Feynman, R. P. (1974). Cargo cult science. Engineering and Science, 37(7), 10-13.

[6] Birhane, A., & van Dijk, J. (2020). Robot Rights? Let's Talk about Human Welfare Instead. arXiv preprint arXiv:2001.05046.

Additionally, we'd like to point out that there are tasks achieved by AI systems today that don't have clear parallels in terms of human intelligence and aren't effectively captured by the proposed definition. As an example, the nudging and shaping of human behaviour on large social media platforms[7] isn't something that human intelligence can be known to exhibit in a traditional sense. Specifically, a lot of it is emergent from both directed efforts by the undergirding platform algorithms to achieve specific goals (say for example, higher clickthrough rates and engagement on content to generate ad revenues[8]) and undirected efforts as humans interact with each other and the platform dynamics to achieve their own goals such as seeking out information, connecting with others, etc.

Speaking just of computers in the definition also has limitations in terms of the perception that is created when thinking about the devices where such intelligence is captured, especially as we move towards embedding this into all sorts of devices that operate in the background, fading into ambient intelligence[9] with the Internet of Things (IoT) as an accelerant for this trend. Thus, the use of the word "computers" may create a false expectation without further clarifying that what is meant by that is "computing devices" which captures other devices as mentioned above which can also exhibit rule-based or intelligent behaviour that should be considered, especially as edge computing[10] becomes more powerful and computations, learning, and inference are not only performed in centralized data centers. Platforms where interactions are mediated by both artificial and human agents[11] also don't strictly fall into this framing of the definition as the "intelligence" is captured in a more distributed sense[12] causing significant shifts in people's behaviour.

From a regulatory standpoint and incorporating this into the definition as it relates to the AI strategy, it will be of paramount importance to articulate these nuances as a supplemental subtext to the primary definition proposed or by revising and expanding the proposed definition such that these expectations are made clear and lead to meaningful and measured strides in investments for AI development and directing regulatory oversight in an appropriate manner.

---

[7] Luceri, L., Braun, T., & Giordano, S. (2019). Analyzing and inferring human real-life behavior through online social networks with social influence deep learning. Applied network science, 4(1), 34.

[8] Theocharous, G., Thomas, P. S., & Ghavamzadeh, M. (2015, June). Personalized ad recommendation systems for life-time value optimization with guarantees. In Twenty-Fourth International Joint Conference on Artificial Intelligence.

[9] Ramos, C., Augusto, J. C., & Shapiro, D. (2008). Ambient intelligence—the next step for artificial intelligence. IEEE Intelligent Systems, 23(2), 15-18.

[10] Zhu, G., Liu, D., Du, Y., You, C., Zhang, J., & Huang, K. (2020). Toward an intelligent edge: wireless communication meets machine learning. IEEE Communications Magazine, 58(1), 19-25.

[11] Ma, X., & Brown, T. W. (2020). AI-Mediated Exchange Theory. arXiv preprint arXiv:2003.02093.

[12] Zeng, D., Chen, H., Lusch, R., & Li, S. H. (2010). Social media analytics and intelligence. IEEE Intelligent Systems, 25(6), 13-16.

## Do you agree that the strategy should be people-centred and aligned with Scotland's National Performance Framework?

The NPF provides a much needed foundation for aligning the AI strategy such that the development and deployment of AI-enabled solutions truly leads to the inclusion of all people in Scotland to partake in economic and social development.

The NPF highlights a lot of ideas that align quite well with the established domains of responsible, inclusive, and ethical development of AI. Specifically, we see this serving as a great opportunity for the drafters of the final strategy to refer to the breadth of literature in refining the concepts of fairness[13], inclusion[14], ethics[15], explainability[16], responsibility[17], transparency[18], accountability[19], and other related concepts from the perspective of the societal impacts of AI.

We also strongly advocate for a participatory approach[20] in the design of the strategy as has been undertaken by the committee in this initial phase for the public consultation. Our experience running workshops for various public entities and government agencies has surfaced the power in seeking diverse perspectives outside of traditional academic circles, industry and other avenues and instead engaging in a long-running, repeated engagement methodology. As an example, quite frequently, there is very little discussion and consultation with on-the-ground workers who might be affected by the use of automation from a labor perspective[21], specifically, there is little deep, informed consultation with such workers to understand their aspirations and what role they see themselves playing in an economy that might have high degrees of automation.

---

[13] Mehrabi, N., Morstatter, F., Saxena, N., Lerman, K., & Galstyan, A. (2019). A survey on bias and fairness in machine learning. arXiv preprint arXiv:1908.09635.
[14] Miller, F. A., Katz, J. H., & Gans, R. (2018). The OD imperative to add inclusion to the algorithms of artificial intelligence. OD PRACTITIONER, 50(1), 8.
[15] Jobin, A., Ienca, M., & Vayena, E. (2019). The global landscape of AI ethics guidelines. Nature Machine Intelligence, 1(9), 389-399.
[16] Adadi, A., & Berrada, M. (2018). Peeking inside the black-box: A survey on Explainable Artificial Intelligence (XAI). IEEE Access, 6, 52138-52160.
[17] Dignum, V. (2017). Responsible artificial intelligence: designing AI for human values.
[18] Wischmeyer, T. (2020). Artificial Intelligence and Transparency: Opening the Black Box. In Regulating Artificial Intelligence (pp. 75-101). Springer, Cham.
[19] Doshi-Velez, F., Kortz, M., Budish, R., Bavitz, C., Gershman, S., O'Brien, D., ... & Wood, A. (2017). Accountability of AI under the law: The role of explanation. arXiv preprint arXiv:1711.01134.
[20] Katell, M., Young, M., Dailey, D., Herman, B., Guetler, V., Tam, A., ... & Krafft, P. M. (2020, January). Toward situated interventions for algorithmic equity: lessons from the field. In Proceedings of the 2020 Conference on Fairness, Accountability, and Transparency (pp. 45-55).
[21] De Stefano, V. (2018). 'Negotiating the Algorithm': Automation, Artificial Intelligence and Labour Protection. Artificial Intelligence and Labour Protection (May 16, 2018). Comparative Labor Law & Policy Journal, Forthcoming.

Often it might also be the case that there is a desire to push back against AI-enabled solutions because of a misalignment from the perspective of local values and other considerations. This can be seen in the case of indigenous populations whose perspectives are often not considered [22] when thinking about investments in AI development and deployment and how they can be included in the benefits that arise from the use of AI. Most of the strategies tout inclusion as an important facet to be considered and we are thrilled with the alignment that Scotland's AI strategy takes with the NPF. We simultaneously would like to emphasize to involve researchers who specialize in demographic analysis and historians who can take a critical lens on the strategy such that traditionally marginalized populations have a chance to voice their aspirations and include them as a part of building AI in Scotland that does indeed bring prosperity to all and trigger benefits for people beyond just the majority. From our work, we've found that such a process does require effective level-setting so that people have a shared vernacular when it comes to both the technical and social science concepts. This is crucial so that interpretations are not lost in translation between diverse audiences.

People-centrism also requires broadening thinking around the second-order effects of AI-enabled solutions. An example comes from how people with disabilities face troubles in interacting with spaces around them that are developed without their needs in mind. We see this become particularly problematic in the case of AI, as an example how online conversations might lose diversity[23] because of biases in AI development that utilize large-scale corpora sourced from the internet for development. A more deliberate approach should be mandated by the strategy document to ensure that such problems are not discovered post-hoc at which point fixing them is not only more costly but leads to a bandage approach.

We also see that true inclusion and development of AI such that it leads to the flourishing of all in a sustainable manner requires the empowerment of local communities such that they can utilize this general-purpose technology to solve for their communities borrowing on insights and intelligence that they have by being the closest to the problem. On that note, democratization of AI-enabled technologies[24], specifically by investments in making larger public datasets available for training of AI systems under the supervised machine learning paradigm, more access to open compute infrastructure that allows from those outside of large corporations and research labs to experiment and develop AI solutions[25], and accessible training to build and use AI-enabled solutions will be necessary in achieving the goals as set out in the NPF. This will allow smaller organizations to thrive as well, creating more opportunities for distributing the gains from using AI within Scotland.

---

[22] Maitra, S. (2020, February). Artificial Intelligence and Indigenous Perspectives: Protecting and Empowering Intelligent Human Beings. In Proceedings of the AAAI/ACM Conference on AI, Ethics, and Society (pp. 320-326).

[23] Hutchinson, B., Prabhakaran, V., Denton, E., Webster, K., Zhong, Y., & Denuyl, S. (2020). Social Biases in NLP Models as Barriers for Persons with Disabilities. arXiv preprint arXiv:2005.00813.

[24] Garvey, C. (2018, April). A framework for evaluating barriers to the democratization of artificial intelligence. In Thirty-Second AAAI Conference on Artificial Intelligence.

[25] Brundage, M., Avin, S., Wang, J., Belfield, H., Krueger, G., Hadfield, G., ... & Maharaj, T. (2020). Toward Trustworthy AI Development: Mechanisms for Supporting Verifiable Claims. arXiv preprint arXiv:2004.07213.

## How do you think AI could benefit Scotland's people, and how do we ensure that the benefits are shared and no-one is left behind?

Looking at the two aspects posed in this question, we see that the benefits to Scotland's people can be split along the lines of economic benefits and societal benefits. From an industry and economic perspective, there are many aspects of agriculture[26] and forestry[27] that can benefit from the use of AI, this has the added advantage of utilizing some of the modern techniques to boost productivity in land-use but also from the labor productivity that is involved in these sectors. When looking at optimizing for sustainable fishery, there is vast literature that takes a systems thinking approach[28] but this can now be supplemented by insights from using AI as a means to better identify fishing vessels and activity[29] which can help prevent illegal activity that can render fishing unsustainable. Automation in this domain can also help to improve the situation in areas where there might be a labor shortage[30] in the space and help in providing insights for example with fish stock reporting and research[31].

In the manufacturing sector, ample opportunities abound in the form of the utilization of AI towards improving the work being done in optoelectronics, software, and life sciences related industries. Especially, in the domain of financial services, there are tremendous opportunities to use the advances in AI for better credit allocation and risk management[32], regulatory compliance work[33] that requires repetitive manual tasks and prone to human errors, among other innovations[34] that can extend the advantages of formal banking to the hitherto "unbanked". This has downstream implications in enhancing equity and economic growth that is distributed more evenly allowing prosperity for all. Yet, the use of AI in this domain is not without ethical risks[35] and concerns and cautious deployment with appropriate safeguards need to be put in place. From an energy perspective, the push to renewables has brought about some novel challenges with grid balancing due to the variability in power generation which can be better balanced by the use of AI-enabled solutions that can forecast both the demand requirements on the grid and measure those against the potential input from renewable energy sources to reduce wasted resources and enhance the stability of the grid[36] while promoting long-term sustainability in energy use. In the oil and gas industry, there is a potential to optimize drilling operations[37] which can help to reduce the carbon footprint of these environmentally-detrimental activities. Simultaneously, it also presents an opportunity for eking out efficiencies in the entire value-chain from exploration to end-use of the extracted resources[38].

---

[26] Lakshmi, V., & Corbett, J. (2020, January). How Artificial Intelligence Improves Agricultural Productivity and Sustainability: A Global Thematic Analysis. In Proceedings of the 53rd Hawaii International Conference on System Sciences.

[27] Kourtz, P. (1990). Artificial intelligence: a new tool for forest management. Canadian Journal of Forest Research, 20(4), 428-437.

[28] Meadows, D. (1998). Systems Thinking. Stakeholders and Decision-Making: Sustainable Development Trough Integrated Water Management. Beijing, China, Lead International Inc, 105-108.

[29] Kourantidou, M. (2019). Artificial intelligence makes fishing more sustainable by tracking illegal activity.

[30] https://medium.com/syncedreview/ai-provides-solutions-for-the-japanese-fishing-industry-9865cc15cc2f

[31] https://digitalterritory.nt.gov.au/action-plan/ai-technology-improve-fish-stock-reporting-research

[32] Pacelli, V., & Azzollini, M. (2011). An artificial neural network approach for credit risk management. Journal of Intelligent Learning Systems and Applications, 3(02), 103.

[33] https://medium.com/district3/the-finance-and-ai-ecosystem-45d614e0a478

[34] Gupta, A. (2018). The History of AI in Finance.

[35] https://blog.re-work.co/ethics-in-ai-finance-industry/

In the government services context, there are potential efficiencies that can be gained from the use of AI in setting taxation and other macroeconomic policies[39] by testing out simulations using reinforcement learning agents as an example. Other government services[40] to the citizens can be optimized as well by using AI tools to reduce wait and processing times for things like immigration. This again has downstream effects of greater participation from the populace by encouraging more meaningful interactions with the government where response times are rapid and the potential for human errors is reduced. Lower wait times and errors will also extend the use of these facilities to larger swathes of people with the same amount of investment and resources from the government perspective.

When thinking about how benefits can be shared more widely and that no one is left behind, we need to take a critical look at how integration of AI for example in the education sector from an early-stage[41] can empower the Scottish people to better understand the implications of AI technology and participate meaningfully in the shaping of technical and policy measures as it relates to the development and deployment of AI technology. Yet, this requires widespread internet access[42] and resources and wherewithal to utilize digital technologies that form the foundation for the later use of AI. This is a great opportunity to call for even deeper penetration of this basic infrastructure which will ultimately enhance the efficacy of the benefits that the Scottish Government would seek to realize for its people through the deployment of AI-enabled solutions. To ensure the deployment of AI doesn't further exacerbate inequities[43], private corporations can also be called upon to make contributions to some of the points identified in the previous section around making large-scale compute and data infrastructure accessible to more people along with potentially helping to train the populace in building up AI talent, something that is scarce[44] when it comes to industrial applications beyond basic understanding. Akin to a Corporate Social Responsibility (CSR) framework, government related subsidies and incentives could be offered to motivate the investment from private corporations to provide this in public interest.

---

[36] https://www.csis.org/analysis/optimizing-indias-electricity-grid-renewables-using-ai-and-machine-learning-applications

[37] Bello, O., Holzmann, J., Yaqoob, T., & Teodoriu, C. (2015). Application of artificial intelligence methods in drilling system design and operations: a review of the state of the art. Journal of Artificial Intelligence and Soft Computing Research, 5(2), 121-139.

[38] https://www.ey.com/en_ro/applying-ai-in-oil-and-gas

[39] Zheng, S., Trott, A., Srinivasa, S., Naik, N., Gruesbeck, M., Parkes, D. C., & Socher, R. (2020). The AI Economist: Improving Equality and Productivity with AI-Driven Tax Policies. arXiv preprint arXiv:2004.13332.

[40] Reis, J., Santo, P. E., & Melão, N. (2019, April). Artificial intelligence in government services: A systematic literature review. In World conference on information systems and technologies (pp. 241-252). Springer, Cham.

[41] https://www.theverge.com/2019/12/18/21027840/online-course-basics-of-ai-finland-free-elements

[42] http://www2.gov.scot/About/Performance/scotPerforms/indicator/internet

[43] Korinek, A., & Stiglitz, J. E. (2017). Artificial intelligence and its implications for income distribution and unemployment (No. w24174). National Bureau of Economic Research.

Furthermore, when considering the existing workforce that has the potential to be displaced[45] as a result of automation created by the deployment of AI technologies, investment in reskilling programs will be crucial. The benefits will be distributed unevenly and over a lumpy landscape if all workers are not given the opportunity to meaningfully transition[46] from their current job roles and functions to new ones that either move them into complementary roles with AI, or into adjacent domains where automation is yet to happen. From an economic policy standpoint, Universal Basic Income (UBI)[47] can also be explored as an option (something that is being experimented with in various forms during the 2020 pandemic) to counter some of the displacement effects of AI, specifically as a mechanism to ease transition from one job role to another. This will also give workers the confidence and financial security to prepare for a future of work that involves continual learning and incorporates them in a gainful manner in the economy of the future.

To assess whether people might be left behind as a consequence of the use of AI, a tool that might prove to be useful is Algorithmic Impact Assessment (AIA)[48,49] which builds on the concepts of Privacy Impact Assessment (PIA) and other such measures extending the analysis to not only technical components but also social and economic impacts taking into account all stakeholders, something that can be done by adopting a systems thinking approach. This will also serve as an opportunity to highlight some of the groups that are traditionally left out and give the Scottish Government an opportunity to integrate them more deeply into the benefits that are to be realized from the use of AI. Borrowing from existing social work in inclusive design from a policy perspective, it will be crucial to work with representatives from various communities to ensure that their voices are incorporated into the development of these technologies and their use to truly bring benefits to all such that no one is left behind.

---

[44] https://www.forbes.com/sites/bernardmarr/2018/06/25/the-ai-skills-crisis-and-how-to-close-the-gap/
[45] Osoba, O. A., & Welser, W. (2017). The risks of artificial intelligence to security and the future of work. RAND.
[46] Illanes, P., Lund, S., Moursched, M., Rutherford, S., & Tyreman, M. (2018). Retraining and reskilling workers in the age of automation. McKinsey Global Institute. Available at https://www. mckinsey. com/featured-insights/future-of-work/retraining-and-reskilling-workers-in-theage-of-automation, accessed, 29.
[47] Straubhaar, T. (2017). On the economics of a universal basic income. Intereconomics, 52(2), 74-80.
[48] Kaminski, M. E., & Malgieri, G. (2019). Algorithmic Impact Assessments under the GDPR: Producing Multi-layered Explanations. Available at SSRN 3456224.
[49] Reisman, D., Schultz, J., Crawford, K., & Whittaker, M. (2018). Algorithmic impact assessments: A practical framework for public agency accountability. AI Now Institute, 1-22.

## What do you think of the proposed overarching vision of the strategy, and the two strategic goals that are proposed to underpin this?

The overarching vision of the strategy aligns quite well with the national outcomes and showcases the focus of the Scottish Government on being inclusive and encouraging growth that is shared by all. While they are great in terms of aspirations, from our experience working on similar strategies in other countries, we've found it pertinent to supplement these with a series of white papers or other policy documents that can expand and provide nuance to the vision as articulated here. The reasons for doing so are that it provides clear guidance for people who will use the vision statements to align their own strategies, yet because such statements can be interpreted differently by people, there is a risk of misalignment that can render harm on the very people that this is supposed to benefit. From several participants in the workshop that was hosted by MAIEI, it was emphasised that there is a need for more concreteness and guidance on how to precisely implement the vision statements into practice. The proposed mechanisms include working hand-in-hand with the industry stakeholders such that they have an intrinsic understanding of the goals. Additionally, working with the regulatory bodies in each of the industries provides an opportunity to integrate existing understanding and expertise of the industry such that alignment occurs in a meaningful manner with minimal friction and disruption.

One of the things that we found to be quite great about the vision statements is that they highlight that AI is to be used to empower people so that they flourish and that *local* organizations can prosper. We firmly believe that there is tremendous value in equipping local actors with the necessary tools and resources so that they can build solutions and address problems. Relying on external expertise and solutions comes with problems of insufficient understanding of the local cultural and contextual realities, something that often creates barriers to adoption and a mismatch between expectations and reality of what the technology is supposed to achieve.

Additionally, the governance[50] surrounding the development and deployment of AI-enabled solutions will be crucial in achieving the strategic goals as identified. This is not only necessary for achieving the stated goals but also enhancing trust from the public in utilizing these tools. In the wider debate during the 2020 pandemic, it is evident that trust will play a critical role in adoption of new technologies such as contact tracing[51].

---

[50] Winfield, A. F., & Jirotka, M. (2018). Ethical governance is essential to building trust in robotics and artificial intelligence systems. Philosophical Transactions of the Royal Society A: Mathematical, Physical and Engineering Sciences, 376(2133), 20180085.

[51] Cho, H., Ippolito, D., & Yu, Y. W. (2020). Contact tracing mobile apps for COVID-19: Privacy considerations and related trade-offs. arXiv preprint arXiv:2003.11511.

## Do you agree with the representation of Scotland's AI ecosystem outlined in the scoping document? Is it missing anything?

As presented in the scoping document, the AI ecosystem captures the large elements that form and shape the space. While there is a mention of how governance of the AI ecosystem needs to wrestle with the cross-geographical nature of tools that might be built elsewhere but utilized locally by the Scottish people, we would like to emphasize that just as this was the case prior to the existence of the GDPR, the adoption of the GDPR ushered in an era where not only were data and privacy rights brought to the forefront of the discussion but also created a big push in terms of incentives for data localization and processing requirements that complied with the European standards. To continue to do business in the region, companies rapidly scrambled to put in place measures that allowed them to operate in a legally compliant manner in this domain.

We see that strong alignment with other major trading partners in terms of policy and governance measures surrounding AI while maintaining the interests of the Scottish people will be key in avoiding scenarios where sacrifices need to be made in terms of the rights of the Scottish people as it relates to the impacts of AI on them, irrespective of where the solutions are developed. We emphasize the need for alignment with the major trading partners as a reasonable measure for adherence by companies who supply AI-enabled solutions because if the requirements posed are too different from those in other places, then it might create high levels of friction curtailing investments and technology transfer which would stifle the availability of benefits to the Scottish people of the technology that emerges from elsewhere.

MAIEI had made a similar proposal in terms of privacy legislation amendments to the Office of the Privacy Commissioner of Canada for the implications of AI on PIPEDA[52]. In our research, analysis, and consultation with diverse stakeholders who participated in the workshops[53] that we hosted on that subject, it was made evident that adoption could be encouraged by reducing friction and minimizing diverse guidelines for companies to have to follow. There is the additional knock-on effect of reducing market competitiveness when there are requirements that are too varied because only larger firms that have additional resources to tailor their offerings for different jurisdictions would then supply solutions thus limiting choices for buyers and crowding out innovation in the face of large incumbents which ultimately goes against the stated vision of bringing the opportunity to participate in the Scottish economy and bring prosperity to all organizations within Scotland.

---

[52] https://montrealethics.ai/response-to-office-of-the-privacy-commissioner-of-canada-consultation-proposals-pertaining-to-amendments-to-pipeda-relative-to-artificial-intelligence/

[53] https://montrealethics.ai/wp-content/uploads/2020/03/FINAL-DOC-Submission-to-OPC-consultation-pages-60-76.pdf

One element that we would like to see added or perhaps broken out and clarified further is the role of the procurement officers who will function as the key intermediaries obtaining solutions built on AI technology produced by various firms and deploy them as a part of the downstream applications that will use those solutions. In some work that MAIEI is doing with researchers from UMass Amherst and OCADU Toronto, we've identified this role to be a key decision maker that determines the success rate of the goals of ethical, responsible, and inclusive use of AI. Specifically, as a part of the vertical elements in the diamond, we would advocate for the inclusion of specific resources that help these decision makers make accurate assessments of the work being done such that they can meaningfully assess the trustworthiness of the AI system[54]. These can include tools that help to assess fairness such as fairlearn[55], Aequitas[56], assess interpretability using tools like LIME[57], visualizing for training data by tools like Facets[58], among others that help the procurement officer make a meaningful and informed decision that aligns with their strategies for achieving stated social outcomes.

In AI development and governance as mentioned in the detailed breakdown of the ecosystem map in the scoping document, we also see the need for public engagement which is specified as a beneficiary on the right-hand side but we see that building solutions that are ethical, safe, and inclusive requires active engagement of the larger public by the listed parties so that there is a deeper understanding of the needs and specificities[59] of the communities that the solutions are looking to serve. Ultimately, our view is that given that AI will be something that affects everybody[60] it is important to inform and equip everybody so that they are able to meaningfully participate in the shaping of technical and policy measures in the development and deployment of AI. Akin to democracy, the efficacy of the stated intentions of democratic rule are only achieved when there is active and engaged participation from the populace that is the subject of that democratic rule; skewed participation where there is a small group of people making decisions unilaterally without consultation with those that will be affected widens inequity and can lead to dissatisfaction and disengagement from the process in the future. With AI, given the perceived technical knowledge barrier to participation, we see a clear need for the Scottish Government to embark on a wide-scale effort that can fold in people from all walks of life into the diamond framework that characterizes the AI ecosystem in Scotland.

---

[54] See 25.
[55] Agarwal, A., Beygelzimer, A., Dudík, M., Langford, J., & Wallach, H. (2018). A reductions approach to fair classification. arXiv preprint arXiv:1803.02453.
[56] Saleiro, P., Kuester, B., Hinkson, L., London, J., Stevens, A., Anisfeld, A., ... & Ghani, R. (2018). Aequitas: A bias and fairness audit toolkit. arXiv preprint arXiv:1811.05577.
[57] Ribeiro, M. T., Singh, S., & Guestrin, C. (2016, August). " Why should i trust you?" Explaining the predictions of any classifier. In Proceedings of the 22nd ACM SIGKDD international conference on knowledge discovery and data mining (pp. 1135-1144).
[58] Wexler, J. (2017). Facets: An open source visualization tool for machine learning training data. Google Open Source Blog.
[59] Morozov, E. (2013). To save everything, click here: The folly of technological solutionism. Public Affairs.
[60] Elshawi, R., Maher, M., & Sakr, S. (2019). Automated machine learning: State-of-the-art and open challenges. arXiv preprint arXiv:1906.02287.

## Do you have any comments on the strategic themes that will be explored in detail?

When thinking about the widespread adoption of AI, as highlighted in a prior section of this report, we would like to reiterate that AI be deployed in appropriate contexts where there are clear benefits as they align with the local culture and context. The reason to emphasize this is that given all the seeming opportunities and excitement in terms of various actors rushing to bring solutions to the market, the actual needs of end-users and their comfort levels with technology might be overlooked.

As an example, when thinking about how technology shapes the environment and socio-technical ecosystem that we all inhabit, it is of paramount importance to consider the second-order effects that impact people whose voices aren't the loudest and can get crowded out by those of the majority. People with disabilities are often victims of such changes[61] especially when non-automated tools are not made available as an alternative. With the GDPR offering data subjects the right to object to automated decision making[62], it is important to consider how alternatives can be made available as automated technology is widely deployed. When considering demographic makeup, often older citizens have a tougher time negotiating emerging technologies[63] and for cases where there might be a high degree of automation surprise, a term borrowed from its application in aviation settings[64], such as in using spoken natural language interfaces in customer support[65] instead of the IVR systems used typically have the potential to confuse users and discourage them from accessing these services.

Some of the questions posed in the "Ethical and Regulatory Frameworks" section of the scoping document provide a great starting point for the deeper discussion of some of these issues. We strongly advocate for a participatory approach to arriving at answers to those and other questions, perhaps also involving people from outside Scotland in the discussions. While we acknowledge that the core members of the group should come from Scotland owing to an emphasis on utilizing local intelligence, we believe that there might be lessons learned from external parties who can share their learnings from having done this work elsewhere and perhaps share best practices that have emerged from the deployment of such frameworks in those contexts. One area as an example that we have often seen neglected is the area of machine learning security[66] which explores how AI-enabled systems open up new attack surfaces and vulnerabilities that need to be addressed in addition to the typical cybersecurity vulnerabilities in software infrastructure. This is a nascent field and novel attacks and defenses are put forth in both academia and applied work. It requires people to be familiar with this subdomain beyond just the traditional roles that people have when speaking about the domain of AI ethics which covers issues like bias, fairness, transparency, etc.

---

[61] See 23.

[62] https://ico.org.uk/for-organisations/guide-to-data-protection/guide-to-the-general-data-protection-regulation-gdpr/automated-decision-making-and-profiling/what-does-the-gdpr-say-about-automated-decision-making-and-profiling/

[63] Vaportzis, E., Giatsi Clausen, M., & Gow, A. J. (2017). Older adults perceptions of technology and barriers to interacting with tablet computers: a focus group study. Frontiers in psychology, 8, 1687.

[64] Dehais, F., Peysakhovich, V., Scannella, S., Fongue, J., & Gateau, T. (2015, April). "Automation Surprise" in Aviation: Real-Time Solutions. In Proceedings of the 33rd annual ACM conference on Human Factors in Computing Systems (pp. 2525-2534).

[65] https://www.nuance.com/omni-channel-customer-engagement/voice-and-ivr.html

A participatory approach[67] to developing the ethics guidelines as highlighted previously would be the most inclusive approach to eliciting the norms and values that are relevant to the communities where such systems are being deployed. While there is an inherent tension in importing existing technologies from outside that might have different cultural and contextual values embedded in them, there is an urgent need to *a priori* articulate what are the norms and values that matter to the Scottish Government and people so that they can be put front and center before there is a bias imposed by the choices on offer from the various vendors. The largest gaps by far that we've seen in our work when embarking on initiatives that seek to put in place ethical and regulatory frameworks for the domain of AI are those that are concerned with the practicality of the recommendations and their operationalization[68]. From an engineering standpoint, guidelines such as "Do no harm" serve as a great overarching theme but their needs to be a high degree of granularity and guidance that is provided to the people who will ultimately be responsible for building and using these systems. Even more prescriptive guidelines that are rooted in the existing AI ethics literature need to have more of a bias to action that allows actual application of these principles in practice. Some of the tools that we highlighted in a previous section serve as a great example for the level of detail that needs to be provided when making recommendations. And we acknowledge that such prescriptions have a tendency to become outdated and would require constant maintenance but we see that the allocation of dedicated resources in maintaining an up-to-date set of guidelines, akin to the NIST cybersecurity framework[69], which are constantly updated with the changing research and development in this space will not only help the AI strategy achieve its goal but also serve as a benchmark for other countries to follow in building effective policy instruments that are actionable rather than theoretical.

On the subject of skills and knowledge, we really appreciate the list of questions as they cover quite comprehensively the issues that would be required for successful integration of AI that fosters inclusive and sustainable growth. One thing that we would add to that list of questions as an area of concern to discuss is the empowerment of people to assess the systems that they obtain externally for value-alignment. This would arise from a cross-disciplinary collaboration that counts within its ranks both technical and social science researchers, especially those who are able to analyze the social impacts of AI within the historical context of Scotland.

---

[66] Biggio, B., & Roli, F. (2018). Wild patterns: Ten years after the rise of adversarial machine learning. Pattern Recognition, 84, 317-331.
[67] See 20.
[68] See 25.
[69] Teodoro, N., Gonçalves, L., & Serrão, C. (2015, August). NIST CyberSecurity Framework Compliance: A Generic Model for Dynamic Assessment and Predictive Requirements. In 2015 IEEE Trustcom/BigDataSE/ISPA (Vol. 1, pp. 418-425). IEEE.

Encouraging data sovereignty[70] and evoking trust through strong data protection and privacy[71] guarantees will evoke the necessary trust for widespread adoption of AI by the public. Integrating existing mechanisms like data trusts[72] as an intermediary in handling the data passing through the AI ecosystem will play a critical role in the uniform application of the stated goals of the AI strategy across the nation. It will also encourage higher levels of accountability through a coordinated and consistent policy management. We also strongly advocate for "baking in, rather than bolting on" these measures, i.e. designing trust and privacy in the system from the get-go rather than inserting them as an afterthought will not only be more efficacious but also more resource-efficient[73]. Patchwork measures have been demonstrated to not work as well and cost orders of magnitude more leaving much to be desired.

A strategy that is dynamic to adapt to the ever-changing landscape of AI development and regulation and is coordinated with efforts from around the world will be important. Establishing partnerships with other organizations, both government and otherwise, that are investing resources in gaining deeper insights that are actionable and practical will be critical. There is also a strong case to be made for bringing on board experts who have applied experience in the space as preliminary analysis indicates that the domain around AI governance and crafting and implementing of AI strategies is becoming increasingly crowded with noise and discerning the signal within that noise will require researchers and practitioners who are finely attuned to the realities of what the cutting-edge in the field is and what to realistically expect from the field in the short- to medium-term. Balancing these concerns with other long-term concerns should also be done cautiously to ensure that there isn't overinvestment in one time horizon[74] compared to the other, especially in the case where there might be gaps in addressing the more pertinent concerns.

## How can confidence in AI as a trusted, responsible and ethical tool be built?

Transparency and accountability as concepts connect to one another. To achieve the goal of building trusted, responsible, and ethical tools to be built, we will need a transparent and systematic process to go from regulation and governance specifics, to specifics of the technology process, and the verifiability of the associated compliance. If the AI systems are known to come from ethically minded researchers and developers, and that information is made public whereby the welfare and needs of the public are made explicit, perhaps even contrasting that with pure profit maximization, it will help to engender higher levels of trust from the public in the solutions that are developed and deployed.

---

[70] Peterson, Z. N., Gondree, M., & Beverly, R. (2011). A position paper on data sovereignty: The importance of geolocating data in the cloud.
[71] Nepal, S., & Pathan, M. (Eds.). (2014). Security, privacy and trust in cloud systems. Springer Berlin Heidelberg.
[72] O'hara, K. (2019). Data trusts: Ethics, architecture and governance for trustworthy data stewardship.
[73] Ashish, A., Steven, F., & Rahul, T. (2008). Estimating Benefits from Investing in Secure Software Development. Build Security In. Carnegie Mellon University.
[74] Prunkl, C., & Whittlestone, J. (2020). Beyond Near-and Long-Term: Towards a Clearer Account of Research Priorities in AI Ethics and Society. arXiv preprint arXiv:2001.04335.

Transparency can be thought of on various levels: on a product level showcasing how the AI technology functions, and on a governance and regulation level highlighting the processes involved, both in expressing how decisions are made and how disputes are resolved. Redress mechanisms[75] will be especially important as decisions are turned over to automated systems. Accountability poses a bit more of a challenge given that many different actors are involved at different stages, especially when it comes to allocating responsibility and the degree to which each of the actors in the development and deployment of this technology[76] should be held accountable[77].

As an example, consider the case of autonomous vehicles, where in the case of an incident, we might have to determine if the accountability lies with the manufacturer, the driver, the passengers, or others? This might be relevant from the perspective of determining insurance payouts as an example. In the beginning stage, perhaps an incident report is created, which might serve as a pre-emptive mechanism. One would have to take into consideration many of the common ethical concerns such as minimization of unwanted bias. There would also be a performance requirement to benchmark results from across the industry to have a certain minimum degree of accuracy. In the middle stage: there might be a need for a body that can serve as the agency to process and arbitrate the claims made by each party. Here, it will be important that the adjudicators are familiar[78] with the latest advancements of the field and are attuned to the capabilities and the limitations of the system to avoid being hoodwinked into being swayed. In the final stages of this process, there needs to be verification of the cause (of the issue), to ensure that it has been addressed, that tests are conducted, and fixes are deployed to improve the safety for the entire ecosystem. This would also ideally be folded into a judicial structure such that dispute resolution can leverage existing mechanisms whereby if there is unsatisfactory resolution at lower levels, they can be moved up to higher bodies that have greater oversight and constitutional power to adjudicate on these disputes. Ideally, there will be a fair distribution of experts who are able to parse through the claims from a technical perspective to avoid being trapped in situations where the right questions are not being asked about the true capabilities and impacts of such systems[79].

---

[75] Crawford, K., & Schultz, J. (2014). Big data and due process: Toward a framework to redress predictive privacy harms. BCL Rev., 55, 93.
[76] Arnold, T., Kasenberg, D., & Scheutz, M. (2017, March). Value Alignment or Misalignment--What Will Keep Systems Accountable?. In Workshops at the Thirty-First AAAI Conference on Artificial Intelligence.
[77] Čerka, P., Grigienė, J., & Sirbikytė, G. (2015). Liability for damages caused by artificial intelligence. Computer Law & Security Review, 31(3), 376-389.
[78] https://themarkup.org/allstates-algorithm/2020/02/25/car-insurance-suckers-list

An interesting point that was brought up as a part of the discussions in the workshop was around how autonomous systems should self-identify and if there should be rules mandating that they make explicit that they are artificial entities rather than natural ones, especially as the capabilities of these systems increase. This is captured in the idea of the Turing Red Flag[80]. We have already seen systems that are quite convincing in their ability to mimic humans, and when such systems don't correctly identify themselves as being artificial, it can lead to ethical concerns in terms of what the human users might expect from the system and if they are manipulated into believing that they are interacting with a human being. As a recent example, the Google Duplex system was able to mimic human ticks in speech that gave the other person on the line the impression that they were interacting with a human when in fact they were interacting with a smart voice assistant when making a reservation[81]. Coming back to the example of autonomous vehicles, having clear Gibsonian affordances will be essential in building trust in these systems[82], especially during a transition phase where they might be interacting with a mixture of human-driven and agent-driven systems.

When considering the opacity in the development of the algorithmic systems, especially as it relates to the use of proprietary code bases and datasets, an important question to be asked is the degree to which the information associated with them needs to be made public. A parallel can be drawn with the tax audit system where there is an agency that is trusted as the public agency that has both the resources and legal backing to investigate individuals and businesses for compliance with the laws of the land. A similar exercise could be done in the public interest for algorithmic systems[83], but this would require a competent authority to be set up that is able to effectively carry out such audits and also have the constitutional backing to bring legal action where there is malfeasance detected. Companies could be, in a similar vein as the disclosure of financial statements for publicly traded companies, compelled to share information in terms of meta-data about the datasets that are used to train their AI systems along with other information that can help the public hold these companies accountable while balancing this disclosure with the company's desire to keep proprietary information private.

As discussed previously, such transparency could also aid in developing and enforcing redressal mechanisms which would be essential in maintaining high levels of trust from the public in these systems. Similar to how restaurants are required to display ratings and audit results from health agencies for hygiene, one could make a similar argument for portions of the algorithmic audits to be made publicly available such that they indicate to the public the degree of data hygiene and adherence to the stated ethical, safety, and inclusivity goals as set out in the AI strategy. Having trusted and prequalified lists of vendors[84] that are able to supply solutions that meet these standards can be something that can be explored from a government perspective. Having public ratings that are made available in a centralized source and inspectable by the general population will also increase the likelihood that the companies will adhere to the stated principles and levels of assurance that they offer in terms of building responsible AI systems.

---

[79] https://www.theatlantic.com/technology/archive/2018/04/3-questions-mark-zuckerberg-hasnt-answered/557720/

[80] Walsh, T. (2016). Turing's red flag. Communications of the ACM, 59(7), 34-37.

[81] O'Neal, A. L. (2019). Is Google Duplex too human?: exploring user perceptions of opaque conversational agents (Doctoral dissertation).

[82] Hulse, L. M., Xie, H., & Galea, E. R. (2018). Perceptions of autonomous vehicles: Relationships with road users, risk, gender and age. Safety Science, 102, 1-13.

[83] Casey, B., Farhangi, A., & Vogl, R. (2019). Rethinking Explainable Machines: The GDPR's Right to Explanation Debate and the Rise of Algorithmic Audits in Enterprise. Berkeley Tech. LJ, 34, 143.

Effective communication will play a critical role in the empowerment of people, one of the gaps identified in the workshop that was hosted by MAIEI was that there was a dire need for translation of documentation around AI systems, including their capabilities and limitations. While neural machine translation systems are certainly making strides in easing the seamless translation of internet content into different languages, it will be essential while the capabilities of these systems ramp up to have documentation for these systems available in local languages that make them accessible to those for whom English might not be their native language.

In furthering the discussion around inclusivity, it is also important to think about certain individuals who require greater protections than other demographics, for example children[85], because of their particular susceptibility to harm from the deployment of digital technologies. As is the case with specialized laws that are enshrined to protect the rights of children[86], one would have to take into consideration how some of the requirements and measures as highlighted above will be able to operate in a model where there are such vulnerable populations involved. This ties into the ideas around informed consent[87] which have been quite challenging to implement into practice. As an example under the GDPR, where there is a requirement to have the legal terms be such that they are accessible and not overly laden with legal jargon, realizing that in practice such that users are truly informed and are able to understand the consequences of agreeing to the terms put forth by a service is still a challenge. Default opt-in needs to be removed such that the user has a choice before being cornered into accepting terms that they don't fully understand. There are studies that show that it is infeasible for a person to read through all the terms and conditions for a service or product that they use, across all services that they use. It takes approximately 76 work days[88] to read through all the policies for the various services that people use over the course of their daily lives. Thus, while putting consent mechanisms in place, it is important that they be mandated to be accessible and understandable to the average user so that they can make informed decisions.

---

[84] https://www.canada.ca/en/government/system/digital-government/modern-emerging-technologies/responsible-use-ai/list-interested-artificial-intelligence-ai-suppliers.html

[85] Dias, P., Brito, R., Ribbens, W., Daniela, L., Rubene, Z., Dreier, M., ... & Chaudron, S. (2016). The role of parents in the engagement of young children with digital technologies: Exploring tensions between rights of access and protection, from 'Gatekeepers' to 'Scaffolders'. Global Studies of Childhood, 6(4), 414-427.

[86] https://www.ftc.gov/enforcement/rules/rulemaking-regulatory-reform-proceedings/childrens-online-privacy-protection-rule

[87] https://ico.org.uk/for-organisations/guide-to-data-protection/guide-to-the-general-data-protection-regulation-gdpr/lawful-basis-for-processing/consent/

## Please comment on any other aspect of AI that you feel it is important for Scotland's AI Strategy to address.

From a data infrastructure perspective, something that should be covered is biometric data and how that will be processed and used in the context of artificial intelligence. Fraud techniques will evolve[89] to take advantage of these immutable characteristics and pose a rising threat that will need to be addressed with techniques that go beyond the data protection standards for other pieces of personally identifiable information (PII).

Another aspect to cover is tackling head-on the inherently stochastic nature of deep learning systems. Especially, as it relates to how humans interact with and interpret the results from these systems when they act as decision aids for further downstream tasks like using recidivism predictions for making parole decisions. The key problem there is to train humans who are using these decision aid systems to understand that there are inherent uncertainties in terms of the predictions made by the system. As an example, people have a hard time interpreting the difference between 66% and 78% probability and while decision thresholds that are captured in the model can be tuned for achieving accuracy, they are also ultimately a function of the choices made by the developers of the system and thus capture the inherent stochasticity of this process. People need to be made explicitly aware to not blindly trust these systems, mathwashing[90] is the term that is used to describe this phenomenon whereby people place a higher degree of trust in numerical systems assuming them to be free of human biases when in fact they are just codifying and quantifying human biases.

As machine learning becomes more accessible through techniques like AutoML[91] that make it easier for people who aren't highly technical to develop and deploy such systems, it is imperative that there are concrete guidelines that they can follow to ensure responsible AI systems are being built. Even upstream tasks like data labelling and collection have many implications and organizations doing first-hand collection of this data should have access to a central resource that details best practices to follow and pitfalls to avoid. There is a lot of normative judgement that can end up being imposed in how the data is labelled[92], which can have particularly stark effects in the supervised machine learning context.

---

[88] https://www.theatlantic.com/technology/archive/2012/03/reading-the-privacy-policies-you-encounter-in-a-year-would-take-76-work-days/253851/

[89] Gupta, A. (2018). The Evolution Of Fraud: Ethical Implications In The Age Of Large-Scale Data Breaches And Widespread Artificial Intelligence Solutions Deployment. International Telecommunication Union Journal, 1, 0-7.

[90] Benenson, F. (2016). Mathwashing. Facebook and the zeitgeist of data worship.

[91] Feurer, M., & Hutter, F. (2018). Towards further automation in automl. In ICML AutoML workshop (p. 13).

Ultimately the strategy must address some of the details that determine the success in achieving the stated goals and vision of the strategy. There are many interpretations when broad statements are used. To ensure that there is a consistent interpretation and application of those principles and ideas which truly incorporate the diversity of Scotland, it will be crucial to integrate and work in tandem with people who will be using these systems in their work. Additionally, inclusivity and total transparency[93] in the process will be essential to meeting the goals of evoking a high degree of trust from the Scottish people, their flourishing, and the sustained prosperity of the organizations in Scotland.

---

[92] Eickhoff, C. (2018, February). Cognitive biases in crowdsourcing. In Proceedings of the eleventh ACM international conference on web search and data mining (pp. 162-170).

[93] Benrimoh, D., Israel, S., Perlman, K., Fratila, R., & Krause, M. (2018, June). Meticulous Transparency—An Evaluation Process for an Agile AI Regulatory Scheme. In International Conference on Industrial, Engineering and Other Applications of Applied Intelligent Systems (pp. 869-880). Springer, Cham.